\documentclass{article}
\usepackage {graphicx}
\begin{document}
\baselineskip .75cm 
\begin{titlepage}
\title{\bf General and Stronger Uncertainty Relation}       
\author{Vishnu M. Bannur  \\
{\it Department of Physics}, \\  
{\it University of Calicut, Kerala-673 635, India.} }   
%\date{}
\maketitle
\begin{abstract}
Recently, Maccone and Pati [Phys. Rev. Lett. {\bf 113}, 260401 (2014)]  derived few inequalities among variances of incompatible operators which they called stronger uncertainty relations, stronger than Heisenberg-Robertson or Schrodinger uncertainty relations. Here we generalize their study to get infinite number of such inequalities and propose that only one of them may be the correct uncertainty relation between incompatible operators. We get back well known uncertainty relations of Heisenberg-Robertson and Schrodinger under certain limits. We also reexamine the conclusions of Maccone and Pati and comment on their work.    
\end{abstract}
\vspace{1cm}
                                                                                
\end{titlepage}
%%%%%%%%%%%%%
\section{Introduction :}

It is interesting to note that even after several years of development of Heisenberg uncertainty relation and quantum mechanics, Maccone and Pati \cite{pa.1} came up with few inequality relations involving variances of incompatible operators which they called uncertainty relations. They argue that the lower bound of the standard uncertainty relations can be null and hence trivial even for incompatible operators. In order to solve this problem they proposed these inequality relations which they claim to have non-trivial bounds and hence they termed them as stronger uncertainty relations. Here we show that their uncertainty relations are few of an infinite number of such inequality relations. Out of these infinite number of inequalities, we derive the true stronger uncertainty relation which goes to standard uncertainty relations under certain limit.  Then we comment and point out some corrections on their conclusions. Subject of uncertainty relations and minimum uncertainty states, and so on, are frequently used in the formulation of quantum mechanics and recently   quantum optics, quantum computations, so on, and this new development may have consequence on them.   

\section{Heisenberg-Robertson and Schrodinger uncertainty relations:} 

According to quantum postulates, the trajectory of particle is not well defined as in classical mechanics. There may be uncertainties in it's trajectory in phase space and hence the uncertainties in the measurement of position and momentum. Heisenberg \cite{he.1} put it in a quantitative expression $\Delta x \Delta p \ge \frac{\hbar}{2}$ which is known as Heisenberg uncertainty relation. Later, Robertson \cite{ro.1} gave an explicit derivation for any two incompatible operators based on Cauchy-Schwartz inequality. Let $|\psi>$ represent the state of a system and $A$, $B$ are two incompatible operators of interest. According to Heisenberg uncertainty relation, it is impossible to measure precisely and simultaneously the observable corresponding to $A$ and $B$. There will be always uncertainties in the observable of $A$ and $B$, defined as $\Delta A ^2 \equiv <A^2> - <A>^2$ and $\Delta B ^2 \equiv <B^2> - <B>^2$, respectively. such that,
\begin{equation}
  \Delta A \Delta B \ge \frac{1}{2}|<[A,B]>| \,\,\, ,
\end{equation}  
where $[A,B]$ is commutator of $A$ and $B$, and $<O>$ is defined as the expectation value $ <\psi|O|\psi>$ for any operator $O$ with respect to the normalized state $|\psi>$. It may be derived by considering two states defined as $|\psi_A> \equiv [A - <A>] |\psi>$ and $|\psi_B> \equiv [B - <B>] |\psi>$ and applying Schwartz inequality, we get, 
\begin{equation}
  |<\psi_A |\psi_B>|^2 \le \parallel \psi_A \parallel ^2 \parallel \psi_B \parallel ^2\,\,\, , \label{eq:sh} 
\end{equation}
where $\parallel \psi_A \parallel ^2$ is the norm-square of $|\psi_A>$ which on simplification leads to $\Delta A ^2$. Similarly  $\parallel \psi_B \parallel ^2 = \Delta B ^2$. Heisenberg-Robertson uncertainty relation may be obtained from above equation by rewriting as,  
\begin{equation}
  \Delta A ^2 \Delta B ^2 \ge  | Im(<\psi_A |\psi_B>)|^2  = \frac{1}{4}|<[A,B]>|^2\,\,\, , \label{eq:hr}
\end{equation}
where $Im$ refers to imaginary part. If we keep both real and imaginary parts of the left hand side (LHS) of Eq. (\ref{eq:sh}), we get Heisenberg-Robertson and Schrodinger (HRS) \cite{sc.1} uncertainty relation,  
\begin{equation}
  \Delta A ^2 \Delta B ^2 \ge  \frac{1}{4}|<[A,B]>|^2 + \frac{1}{4} |<\{A,B\}> - 2 <A> <B>|^2,
\end{equation}
where $\{A,B\}$ is anti-commutator. 
%It may be stronger than Heisenberg uncertainty relation. 

\section{More general uncertainty relations:} 

We are familiar with uncertainty relations $\Delta x$ $\Delta p$ $\ge \frac{1}{2}|<[x,p]>| = \frac{\hbar  }{2} $ and $\Delta J_i \Delta J_j \ge \frac{1}{2}|<[J_i,J_j]>| = \frac{\hbar  }{2} \epsilon_{ijk} |<J_k>|$, etc., which are all Heisenberg-Robertson uncertainty relations. $x$ and $p$  are position and momentum operators which are incompatible operators, and $J_i$ with $i=x,y,z$ are components of angular momentum operators which are also incompatible operators. Uncertainties $\Delta x$, $\Delta p$ and $\Delta J_i$ are observed on making a measurement on a similarly prepared (ensemble) system., say, in the state $|\psi>$. Expectation values are also with respect to $|\psi>$. $|\psi>$ may be pure state or mixed state. As we discussed in the earlier section, HRS uncertainty relation is derived from Schwartz inequality $  |<\psi_A |\psi_B>|^2 \le \parallel \psi_A \parallel ^2 \parallel \psi_B \parallel ^2$. Both $|\psi_A>$ and $|\psi_B>$ are orthogonal to $|\psi>$. We may generalize it by replacing $|\psi_B>$ by an arbitrary normalized state orthogonal to  $|\psi>$, say, $|\psi^{\perp }>$ and  $|\psi_A>$ by $|\phi> \equiv |\psi_A> + i \alpha |\psi_B>$ with an arbitrary real parameter $\alpha$ \cite{rn.1}, such that the Schwartz inequality now reads as $  |<\phi|\psi^{\perp }>|^2 \le \parallel \phi \parallel ^2$, which on simplification leads to 
\begin{equation}
  \left[ \Delta A^2 - |<\psi^{\perp} |\psi_A>|^2 \right]  + \alpha^2 \left[ \Delta B^2 - |<\psi^{\perp} |\psi_B>|^2\right] + i \alpha \left[\{  <\psi_A |\psi_B> - <\psi^{\perp} |\psi_B>   <\psi_A|\psi^{\perp}>\}   - \{  c.c\}  \right] \geq 0\,\,,  \label{eq:al} 
\end{equation}
where $\{c.c\}$ means complex conjugate of previous terms inside the square bracket. Scalar products $<\psi^{\perp}|\psi_A>$ and $<\psi^{\perp}|\psi_B>$may be simplified to $<\psi^{\perp}|A|\psi> $ and $<\psi^{\perp}|B|\psi>$, respectively. $\alpha$  is a free parameter. This is an inequality relation involving the variance of incompatible operators and we can get infinite number of relations depending on infinite number of values for $\alpha$. In the next section, we will point out that the inequality relations derived by Maccone and Pati \cite{pa.1} may be obtained from above equation for few specific values of $\alpha$ and above equation is more general. However, we argue that the true uncertainty relations may be obtained by fixing $\alpha$ on minimizing above expression and we get, 
\begin{equation}
  \alpha = - \frac{i}{2} \frac{\left[\{  <\psi_A |\psi_B> - <\psi^{\perp} |\psi_B>   <\psi_A|\psi^{\perp}>\}   - \{  c.c\}  \right]}{\left[ \Delta B^2 - |<\psi^{\perp} |\psi_B>|^2\right]} \,\,\, ,
\end{equation}
and inequality Eq.(\ref{eq:al}) reduces to  
\begin{equation}
  \left[ \Delta A^2 - |<\psi^{\perp} |\psi_A>|^2\right] \left[ \Delta B^2 - |<\psi^{\perp} |\psi_B>|^2\right] \geq   
  \frac{1}{4} \left|  <[A,B]> - [\{ <\psi^{\perp} |\psi_B>   <\psi_A\psi^{\perp}>\}  - \{ c.c\} ]   \right|^2 \,\,, \label{eq:n1}
\end{equation}
which is just the modified or generalized uncertainty relation of  Heisenberg and Robertson, Eq.(\ref{eq:hr}).
It may be convenient to reduce it to another inequality,
\begin{equation}
   \Delta A^2 + \Delta B^2 \geq   
  \left|  <[A,B]> - [\{ <\psi^{\perp} |\psi_B>   <\psi_A|\psi^{\perp}>\}  - \{ c.c\} ]   \right|   + |<\psi^{\perp} |\psi_A>|^2 + |<\psi^{\perp} |\psi_B>|^2  \,\,,
\end{equation}  
so that we can get the expression for the sum of variances. 

Our choice of $|\phi> = |\psi_A> + i \alpha |\psi_B>$ to derive above equations is still not general. We may choose $|\phi> = |\psi_A> + (\beta + i \alpha) |\psi_B>$ with two free parameters and again proceeding the same way, we get, 
\[
 \left[ \Delta A^2 - |<\psi^{\perp} |\psi_A>|^2\right] \left[ \Delta B^2 - |<\psi^{\perp} |\psi_B>|^2\right] \geq    \frac{1}{4} \left|  <[A,B]> - [\{ <\psi^{\perp} |\psi_B>   <\psi_A|\psi^{\perp}>\}  - \{ c.c\} ]   \right|^2
 \]
\begin{equation}
  + \frac{1}{4} \left|  <\{A,B\}> - 2 <A> <B> - [\{ <\psi^{\perp} |\psi_B>   <\psi_A|\psi^{\perp}>\}  + \{ c.c\} ] \right|^2 \,\,\, , \label{eq:gh}
\end{equation} 
which is the generalization of Heisenberg-Robertson and Schrodinger uncertainty relation. It may be also put in the form,
\[
 \Delta A^2  + \Delta B^2 \geq |<\psi^{\perp} |\psi_A>|^2 + |<\psi^{\perp} |\psi_B>|^2 + \mbox{\huge $[$}
 \left|  <[A,B]> - [\{ <\psi^{\perp} |\psi_B>   <\psi_A|\psi^{\perp}>\}  - \{ c.c\} ]   \right|^2 
 \]
\begin{equation}
+ \left|  <\{A,B\}> - 2 <A> <B> - [\{ <\psi^{\perp} |\psi_B>   <\psi_A|\psi^{\perp}>\}  + \{ c.c\} ] \right|^2 \,\, \mbox{\huge $]$} ^{1/2} 
 \,\,\, .  \label{eq:f} 
\end{equation} 
In this equation, we know $|\psi>$, the prepared state of the system which may be pure or mixed state and hence the inequality depends on $|\psi^{\perp}>$. We can think of different choices of $|\psi^{\perp}>$. If it is a null vector, above equations reduce to HRS relations. Suppose $|\psi^{\perp}>$ is $\frac{|\psi_B>}{\Delta B}$, then $<\psi^{\perp} |\psi_B> = \Delta B$, $<\psi^{\perp} |\psi_A> = \frac{<\psi_B|\psi_A>}{\Delta B}$ and hence above equation reduces $\Delta A \Delta B \ge |<\psi_B|\psi_A>|$ which is same as Eq.(\ref{eq:sh}).  Hence we get back the HRS uncertainty relation for the choice of $|\psi^{\perp}>$ equal to $\frac{|\psi_B>}{\Delta B}$ or $\frac{|\psi_A>}{\Delta A}$. Next section we discuss another example of Maccone and Pati in detail.   

\section{Special cases of general inequality Eq.(\ref{eq:al}):} 

Next, we discuss the recent work of Maccone and Pati \cite{pa.1} which motivated us to look into our generalized uncertainty relation. It is a special case of above general formalism for a particular values of parameters, $\beta = 0$ and $\alpha = \mp 1$. That is, from Eq.(\ref{eq:al}) for $\alpha = \mp 1$, we get, 
 \begin{equation}
  \left[ \Delta A^2 - |<\psi^{\perp} |\psi_A>|^2 \right]  +  \left[ \Delta B^2 - |<\psi^{\perp} |\psi_B>|^2\right] \mp i \left[\{  <\psi_A |\psi_B> - <\psi^{\perp} |\psi_B>   <\psi_A|\psi^{\perp}>\}   - \{  c.c\}  \right] \geq 0\,\,, 
\end{equation}
which is same Eq.(3) of Ref. \cite{pa.1},
 \begin{equation}
   \Delta A^2 + \Delta B^2 \geq   
  \pm i <[A,B]>  + |<\psi| A \pm i B|\psi^{\perp}>|^2   \,\,, \label{eq:mp} 
\end{equation}  
after some simplifications. We like to point out that this equation also reduces to HRS relations for the choice of $|\psi^{\perp}>$ equal to $\frac{|\psi_B>}{\Delta B}$ or $\frac{|\psi_A>}{\Delta A}$ which has null or trivial bounds as we discussed earlier. This is in contrary to the conclusions of Maccone and Pati where they choose these  $|\psi^{\perp}>$ as an illustration to prove that the lower bound of their Eq.(\ref{eq:mp}) is nonzero for almost any choice of $|\psi^{\perp}>$. 

The motivation of authors of Ref. \cite{pa.1} to look for a stronger uncertainty relation was the following. When we have $|\psi>$ to be pure state which may be eigen state of  one of the incompatible operators, RHS and LHS of HRS uncertainty relations both goes to zero. So the lower bound of the inequality is null or trivial. For example, in the case of angular momentum operators, say, $\Delta J_z \Delta J_x \ge \frac{1}{2}|<[J_z,J_x]>|$, both sides are zero for $|\psi> = |jm>$, angular momentum state such that $J_z |jm> = m \hbar |jm>$.  This is the reason to formulate uncertainty relations with nontrivial inequality, or stronger uncertainty relations. But note that if our system is in the state $|jm>$, then there will not be uncertainty in $J_z$ ($\Delta J_z = 0$). There may be uncertainty in other components which are related by $\Delta J_x \Delta J_y \ge \frac{1}{2}|<i \hbar  J_z>|$,  which is not trivial in general. Similarly, the uncertainty relation $\Delta x$ $\Delta p$ $\ge \frac{1}{2}|<[x,p]>| = \frac{\hbar  }{2} $  relation is also an exceptional case and both sides are nontrivial.

Another example we may consider is the spin $1$ particle state, discussed in Ref.\cite{pa.1}, $|\psi> = \cos \theta |+> + \sin \theta |->$  and $|\psi^{\perp}>$ may be taken as $|0>$. Let $A=J_x$ and $B = J_y$ so that $[A,B] = i \hbar J_z$. Therefore, for this state $<[A,B]>$ may be evaluated as $i \hbar ^2 \cos 2 \theta$ and hence HRS uncertainty relation gives $\Delta J_x \Delta J_y \ge \frac{\hbar ^2}{2} \cos 2 \theta $ or $\Delta J_x^2 + \Delta J_y^2 \geq \hbar ^2 \cos 2 \theta $ . Matrix element   $<\psi|A \pm i B|0> = <\psi|J_x \pm  i J_y|0>$ gives $\sqrt{2} \hbar \cos \theta$ and $\sqrt{2} \hbar \sin \theta$ for $+$ and $-$ respectively. It immediately follows that inequality Eq.(\ref{eq:mp}) is non-trivial for all $\theta$ and  we get $\Delta J_x^2 + \Delta J_y^2 \geq \hbar ^2 $ for both the values of $\alpha = \mp 1$ . We can also evaluate $\Delta A^2 = \Delta J_x^2 = \hbar ^2 (1 + \sin 2 \theta )$ and $\Delta B^2 = \Delta J_y^2 = \hbar ^2 (1 - \sin 2 \theta )$ and hence it corresponds to equality in Eq.(\ref{eq:mp}). 

Let us consider the same example of spin 1 states and apply our general uncertainty relation Eq.(\ref{eq:f}). We evaluate $<\psi^{\perp}|A|\psi> = <0|J_x|\psi> = \frac{\hbar}{\sqrt{2}} (\sin \theta + \cos \theta)$ and $<\psi^{\perp}|B|\psi> = <0|J_y|\psi> = \frac{i \hbar}{\sqrt{2}} (\cos \theta - \sin \theta)$ and RHS side of the Eq.(\ref{eq:f}) reduces to $\hbar ^2$. Again the lower bound of the inequality is nontrivial and equal to $\hbar ^2$ as in the case of Eq.(\ref{eq:mp}). At the same time, product uncertainty relation Eq.(\ref{eq:n1}) still has trivial lower bound, but leads to separate uncertainty relations $\Delta J_x^2 \ge \frac{\hbar^2}{2} (1+ \sin 2 \theta)$ and $\Delta J_x^2 \ge \frac{\hbar^2}{2} (1- \sin 2 \theta)$. It implies that $\Delta J_x \Delta J_y \ge \frac{\hbar^2}{2} \cos 2 \theta$ which coincides with result of HRS uncertainty relation which has nontrivial bounds in this case as pointed out earlier.

Now we consider the same problem of spin 1 system, but interchanging the states. That is, let us take $|\psi> = |0>$ which is the state of the system and $|\psi^{\perp}> = \cos \theta |+> + \sin \theta |->$ so that we can have infinite number of orthogonal states depending on the values of $\theta$. Now the general uncertainty relation Eq.(\ref{eq:f}) gives $\Delta J_x^2 + \Delta J_y^2 \geq 2 \hbar ^2 \cos ^2 \theta$ and inequality of Ref. \cite{pa.1}, Eq.(\ref{eq:mp}), gives $\Delta J_x^2 + \Delta J_y^2 \geq$ $ 2 \hbar ^2 \cos ^2 \theta$ and $2 \hbar ^2 \sin ^2 \theta$ for $\alpha$ equal to $+1$ and $-1$ respectively. Both have nontrivial lower bounds and at the same time HRS has trivial lower bound since $<0|[J_x,J_y]|0> = 0$. But the modified Heisenberg equation, Eq.(\ref{eq:gh}), leads to, 
 \begin{equation}
  \left[ \Delta J_x^2 - \frac{\hbar^2}{2} (1+ \sin 2 \theta) \right] \left[ \Delta J_y^2 - \frac{\hbar^2}{2} (1- \sin 2 \theta) \right] \geq   
  \frac{1}{4} \hbar^4 \cos ^2 2 \theta  \,\,, 
\end{equation} 
which is consistent with the results of Eq.(\ref{eq:f}). Note that in this example only one solution of Eq.(\ref{eq:mp}) ($\alpha = +1$) coincides with the general uncertainty relation and result depends on $\alpha$. We can also evaluate the variances as $\Delta J_x^2 = \Delta J_y^2 = \hbar^2$ and hence confirms the   inequalities of Eq.(\ref{eq:f}) and Eq.(\ref{eq:mp}). 
  
\section{Conclusions:} 

We generalize the recent work of Maccone and Pati \cite{pa.1} on the stronger uncertainty relation and found that we can have infinite number of such inequalities. Uncertainty relations of Maccone and Pati are just few special cases of such inequalities and need not be the actual uncertainty relations. We derived the correct uncertainty relation by minimizing the inequalities. This true, generalized, stronger uncertainty relation goes to Heisenberg-Robertson and Schrodinger uncertainty relations at different limits. We also verified that the inequality is nontrivial and hence stronger uncertainty relation, using spin-1 system as an example.

\end{document}